# A MULTI MEGAWATT RING CYCLOTRON TO SEARCH FOR CP VIOLATION IN THE NEUTRINO SECTOR

A. Calanna, CSFN-SM, Catania, Italy

L. Calabretta, M. Maggiore, L. A. C. Piazza, D. Rifuggiato, INFN-LNS, Catania, Italy

*Abstract*

A new approach to search for CP violation in the neutrino sector [1,2] is proposed by the experiment called DAEδALUS (Decay At rest Experiment for $δ_{cp}$ At Laboratory for Underground Science). DAEδALUS needs three sources of neutrino fluxes, each one located at 1.5, 8 and 20 km from the underground detector. Here we present the study for a Superconducting Ring Cyclotron able to accelerate the H2+ molecules and to deliver proton beam with maximum energy of 800 MeV and the required high power. The magnetic field produced by the proposed superconducting magnetic sector, simulated by the code TOSCA, the isochronous magnetic field, some preliminary feature on the beam dynamic and the magnetic forces acting on the coils are here presented.

## INTRODUCTION

A new experiment called DAEδALUS (Decay At rest Experiment for $δ_{cp}$ At Laboratory for Underground Science) to search for CP violation in the neutrino sector [1,2] has recently proposed. This experiment needs three neutrino sources produced by a proton beam with energy of about 800 MeV. The nearest site, located at 1.5 km from the underground detector, needs a proton beam with a power of 1 MW. The second source should stay at a distance of about 8 km from the detector and a beam power of 2 MW is required. The third neutrino source, 20 km far from the detector, has to be fed with a proton beam of power higher than 5 MW. An additional constraints requested to the accelerators is to operate with a duty cycle of 20%, typically 100 ms. beam on, 400 ms. beam off properly synchronised. Due to this low duty cycle the peak power is 5 time higher than the average power. Accelerator complex consisting of two cyclotrons, one injector cyclotron and a main ring cyclotron booster, have already proposed as drivers for energy amplifier or waste transmutation plants [3,4]. According to a previous proposal [4,5], a superconducting ring cyclotron able to accelerate a beam of H2+ up to 800 MeV/n with a peak current of 5 mA of H2+, peak power 8 MW, average power 2 MW, is here presented.

The layout of the proposed Superconducting Ring Cyclotron (SRC) is shown in Fig. 1. The injection trajectory with the four magnetic channels (M.C.) and the electrostatic deflector (E.D.) necessary to perform the proper beam injection are shown. The extraction trajectories with energy spread of ±1% are also shown. Only one M.C. is necessary to steer a little the trajectory and to axially focus the beam.

The main parameters of SRC are presented in table 1.

The acceleration of H2+ molecule allows extracting the beam by a stripper foil which breaks the molecules and produces two free protons. Extraction by stripping does not require well separated turns at the extraction radius and allows using lower energy gain per turn during the acceleration process and/or lower radius for the magnetic sectors, with a significant reduction of power losses for the RF cavities and construction cost. The extraction by stripper allows to extract beams with large energy spread. In Fig.1. the extraction trajectories for beam with energy spread of ± 1% are indistinguishable. The energy spread produced by space charge effect on the longitudinal size of the beam is not crucial in this kind of accelerator, and flattopping cavities are unnecessary. Looking at the Fig. 1, it is evident that the most critical part is the inner region which is mainly filled by the RF cavities protrusion. The insertion of two double gap cavities is proposed mainly to increase the beam orbit separation at the injection and to reduce the beam losses due to the interaction of the H2+ beam with the residual gases.

## THE MAGNET SECTOR

The isochronous field is produced by 8 magnet sectors; each sector is powered by a couple of superconducting coils. The coils shape is shown in Fig. 2. Each coil, NbTi made, is divided in 9 parts. For sake of simplicity the coil parts 3, 4, 7, 9 have assumed straights, but in final design they have to be bent to optimize the stress distribution. The coils have been tilted of ±3° respect to the median plane. The minimum and maximum distance between the coils being 15 cm and 60 cm at radii 5.48 m (part 9) and

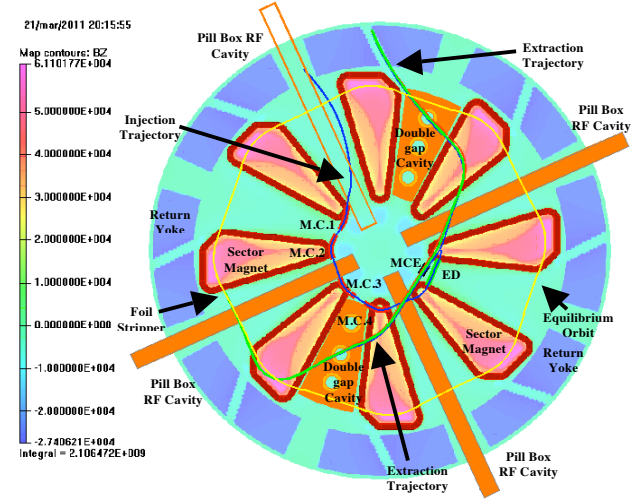

Figure 1: Layout of the Superconducting Ring Cyclotron. The injection and extraction trajectory, with their magnetic channel (M.C.) and the electrostatic deflector (E.D.) for the injection are also shown.

Table 1: Main parameters of the proposed SRC

| $E_{max}$ | 800 MeV/n | $E_{ini}$ | 50 MeV/n |
|---|---|---|---|
| $R_{ext}$ | 4.90 m | $R_{inj}$ | 1.8 m |
| <B> at $R_{ext.}$ | 1.88 T | <B> at $R_{in}$ | 1.06 T |
| Bmax | <6.3 T | Flutter | 1.4÷1.97 |
| Outer radius | ≤7 m | Pole gap | 60 mm |
| Sector height | 5.6 m | N. Sectors | 8 |
| Sector weight | < 500 tons | N. Cavities | 6 |
| 4 Cavities | Single gap | 2 Cavities λ/2 | Double gap |
| Acc. Voltage | 550÷1000 kV | Acc. Voltage | 200÷250 kV |
| Power/cavity | 350 kW | Power/cavity | 300 kW |
| RF | 49.2 MHz | harmonic | 6th |
| <ΔE/turn> | 3.6 MeV | Number of turn | 420 |
| ΔR at $R_{ext}$ | 5 mm | ΔR at $R_{inj}$ | > 10 mm |
| Coil size | 17 x 27 cm$^2$ | Icoil | 5000 A/cm$^2$ |

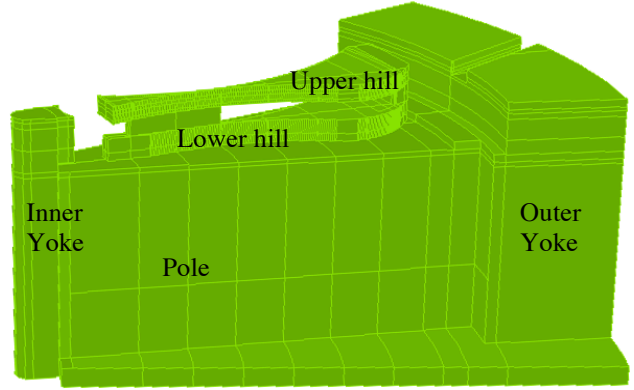

Figure 2: Pair of superconducting coils, tilted of ±3° vs. the median plane. Each coil is divided in 9 parts.

1.5 m (parts 1, 2) respectively. This solution is useful to produce higher field at outer radii and lower field at the inner radii. The main parameters of the coils and forces which act on each part are presented in table 2. The components of the forces are along the axis x, y and z of a canonical frame system with origin in the median plane and at centre of cyclotron. The x axis is direct perpendicular at part 9 of each coil. The z axis is perpendicular at cyclotron median plane. The highest forces act on the longest parts 3 and 4 of the coil and have opposite directions which tend to enlarge the coil. To balance these huge forces and to reinforce the structure of the liquid helium [LHe] vessel we use a connecting plate (2.2 m long and 9 cm thick) to join together the parts 3 and 4 of the coil, this solution has already used in the Riken superconducting ring cyclotron [6]. The connecting plate will be part of the LHe vessel. To have a robust LHe vessel a thickness of 4 cm for the inner wall and for the plate near the median plane was assumed, while for the outer wall and plate far from the median plane a thickness of 7 cm was fixed. This structure is enough strong to support also the vertical attractive force between the pair of coils. The LHe vessels of the upper and lower coils will be connected along the parts 1, 2, 6, 8 and 9 to sustain the huge attractive force. The average pressure on these parts will be less than 1.3 kg/mm$^2$. Unfortunately the total residual forces Fx and Fy are unbalanced and then they have to be supported by a tie rods set that maintain the LHe vessel in the right position. Despite the fact that the configuration here presented shows large values for both Fx and Fy, during our process of optimization we have learned how to minimize both these forces putting additional iron mass near the inner or the outer yoke and near the part 7 of the coil. Anyway, residual component forces also after the optimization procedure to minimise these force it is expected. In particular the Fx has to be compensated by a radial support like in the Riken SRC [6]. The cryogenic insulation is guaranteed by a vacuum chamber with a 3.5 cm gap between the LHe vessel and the room temperature of the cryostat. This vacuum chamber can be filled with more than 15 insulation layers to minimize the thermal losses. The wall thickness of the cryostat is supposed to be 1.5 cm tick. The hill is divided in two, the upper and lower hill, see Fig. 3. Between the two parts there is a large empty volume which will allow the insertion of the connecting plate. This empty volume can be used also to install room temperature trim coils, useful for fine tuning of the magnetic field. From radius 4.6 to 5 m the sector has a spiral shape to increase the vertical focusing property of the machine. The shape of the coil, shown in Fig. 2, was designed just to wrap around the hill and to avoid a concave shape. To achieve the required isochronous field we need a strong variation of the mean field vs. radius, from 3 T, at injection radii, to about 6 T at outer radii of the pole. The modulation of the angular width of the iron pole is not enough to produce the request variation. For this reason the coils had been tilted. This solution offer some additional advantages for the part 1 and 2 of the coil: the maximum magnetic field value is 6.3 T and there is a useful repulsive vertical force to balance the attractive force of parts 3 and 4.

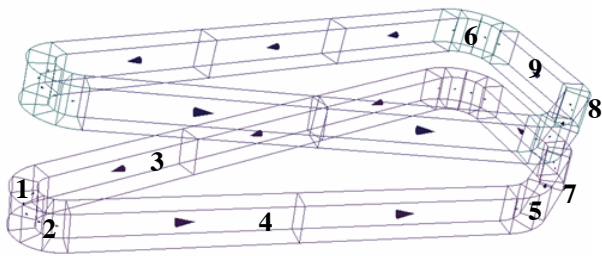

Figure 3: Lower half part of one sector, angular width 45°

Table 2: Main parameters of the superconducting coils

| Coil size | 17 x 27 cm$^2$ | | Icoil | 5000 A/cm$^2$ |
|---|---|---|---|---|
| Max. Field | 6.3 T | | Max. Force | 4.4 ton/cm |
| Coil Part | Length [cm] | Fx [ton] | Fy [ton] | Fz [ton] |
| 1 | 32 | -95 | -69 | -23 |
| 2 | 32 | -99 | -86 | -24 |
| 3 | 340 | -350 | 1.164 | 96 |
| 4 | 290 | -127 | 1.081 | 75 |
| 5 | 21 | 33 | 104 | 34 |
| 6 | 43 | 109 | 151 | 75 |

| 7 | 49 | 145 | 157 | 76 |
| 8 | 18 | 64 | 28 | 31 |
| 9 | 88 | 390 | 0 | 177 |

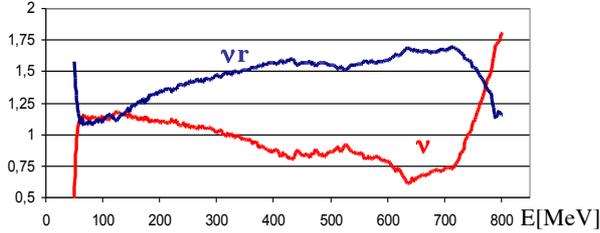

Figure 4: Radial and axial focusing frequencies vs. energy

## BEAM DYNAMIC

The iron of the hill was shaped to match the required ideal isochronous field with an accuracy of ±1%. Although this value is not enough to guarantee the acceleration of the beam, we expect that the further small corrections on the magnetic field, to achieve the good isochronisms with precision better than $10^{-4}$, do not change significantly the focusing properties of the field. A serious effort was necessary to maintain the $\nu z$ higher than 0.5. In fig.4 both the $\nu z$ and $\nu r$ vs. energy are presented. It is quite satisfactory. Despite some resonances are crossed this is done quite fast and we do not expect serious problem.

The beam envelopes for the injection and extraction trajectories are presented in Fig. 5 and 6. The normalized emittance of the beam envelopes shown in both Fig. 5 and 6 is 3.35 π mm. mrad. This large value, about 30 times the emittance of the beam delivered by the ion source [7], takes account of non linear effect along the injection, acceleration and extraction from the cyclotron injector and of space charge effects which produce a serious broadening of the beam emittance. Despite the large emittance the beam envelope is quite small in both the radial and axial plane. Also the energy dispersion along the extraction path is acceptable.

## CONCLUSION

The present study demonstrates that is possible to achieve the required isochronous field with good focusing property, some parameters should be modified to achieve a more reliable solution, however. In particular we would like to reduce the current density from 50 to 42 A/mm$^2$, to reduce the maximum field in the part 1 and 2 of the coil at value below 6 T, to reduce the angular width of the cryostat in the inner region to have more room for the RF cavities and to reduce the magnetic field produced by the coils in the radial range between 2.3 and 3.3 m to have more iron in the hill. To achieve these goals we plan to modify the shape of the coil. The part 6, 8 and 9 of the coils will be parallel to the median plane and with a separation between upper and lower coil of 16 cm. The coils will be tilted from radius 5.18 m to increase the distance from the median plane up to the maximum of 23 cm at radius of 3.3 m, and then stay fixed at this value until the inner radii. To reduce the current density at lower value than 42 A/mm$^2$, we could increase the coil size. This is a problem only at the inner radii, so we need a technical solution to have a coil with a cross section of 20 x 28 cm$^2$ at radii larger than 2.3 m and 14 x 40 cm$^2$ at the inner radii and in particular in the part 1 and 2 of the coil. This solution, if feasible, should reduce the maximum field on the coil's part 1 and 2 and at the same time save room for the RF cavities.

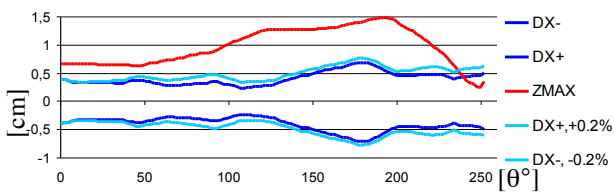

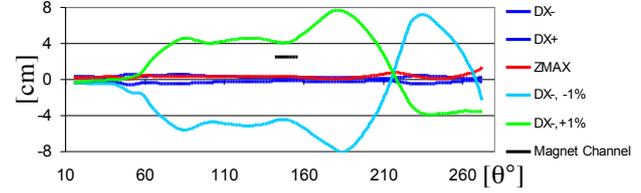

Figure 5: Beam envelope in radial and axial plane along the injection trajectory. Effects due to the energy spread (±0.2%) are negligible.

Figure 6: Beam envelope in radial and axial plane along the extraction trajectory. Effects due to the energy spread (±1%) are acceptable.

It is a pleasure to acknowledge that A. Calanna is supported through funding from MIT.